*Title:* Digital holographic microscopy for the evaluation of human sperm structure.

*Running title*: **Human sperm holography**

*Authors*: **Coppola G[1]., Di Caprio G[2]., Wilding M[1]., Ferraro P[3]., Esposito G[1]., Di Matteo L[1]., Dale R[1]., Coppola G[2]., Dale B[1].**

*Addresses:*

[1]Centro Fecondazione Assistita (CFA-Italia), Via Manzoni 15, 80123 Naples, Italy; [2]Istituto per la Microelettronica e i Microsistemi, del Consiglio Nazionale delle Ricerche, Sezione di Napoli, Via P. Castellino, 111, 80131 Naples, Italy; [3]Istituto Nazionale di Ottica, del Consiglio Nazionale delle Ricerche, Sezione di Napoli, c/o Comprensorio Olivetti, Via Campi Flegrei 34 - 80078 Pozzuoli, Italy.

**Correspondance:** Tel: +39 081641689; Fax: +39 0815479251; e-mail gcoppola@cfa-italia.com

**ABSTRACT**

The morphology of the sperm head has often been correlated with the outcome of in vitro fertilization (IVF), and has been shown to be the sole parameter in semen of value in predicting the success of intracytoplasmic sperm injection (ICSI) and intracytoplasmic morphologically selected sperm injection (IMSI).

In this paper, we have studied whether Digital Holographic (DH) microscopy may be useful to obtain quantitative data on human sperm head structure and compared this technique to high power digitally enhanced Nomarski microscope. The main advantage of DH is that a high resolution 3-D quantitative sample imaging may be obtained thorugh numerical refocusing at different object planes without any mechanical scanning. We show that DH can furnish useful information on the dimensions and structure of human spermatozoo, that cannot be revealed by conventional phase contrast microscopy. In fact, in this paper DH has been used to evaluate volume and indicate precise location of vacuoles, thus suggesting its use as an additional useful prognostic quantitative tool in assisted reproduction technology (ART).

# INTRODUCTION

Following the advent of human in vitro fertilization (Steptoe and Edwards, 1978), much attention has been given to identifying, first embryo morphology, and later, oocyte morphology, as a prognostic tool (Elder and Dale, 2011); less attention has been given to sperm morphology. The spermatozoon delivers the haploid male genome to the oocyte, introduces the centrosome and triggers the oocyte egg into activity.

The sperm head may be considered in three parts; the nucleus with a haploid set of chromosomes, in which deoxyribonucleic acid (DNA) is packaged into a volume that is typically less than 10% of the volume of a somatic cell nucleus (Dadoune, 2003; Elder and Dale, 2011); the acrosome, a large Golgi-derived secretory vesicle on the proximal hemisphere of the head containing an array of hydrolytic enzymes used for digesting the zona pellucida during penetration (Gerton, 2002; Yoshinaga and Toshimori, 2003); and the perinuclear theca, a rigid capsule composed of disulfide bond stabilized structural proteins amalgamated with various other protein molecules (Oko, 1995).

Human spermatozoa exhibit a wide range of shapes. A number of studies indicate that sperm morphology best predicts the outcome of natural fertilization (Kruger et al., 1988; Bartoov et al., 1999), intra-uterine insemination (Berkovitz et al., 1999) conventional IVF (Kruger et al., 1988; Mashiach et al., 1992) and ICSI (Palermo et al., 1993; Bartoov et al., 2003), and several techniques have been described that provide valuable information on the morphology and  pathological features of spermatozoa. In a classical clinical evaluation, human sperm are fixed, stained and analyzed by optical microscopy. Recently, a number of novel techniques have been developed for the identification of more detailed features of cells. Differential interferometric contrast microscopy, scanning near-field optical microscopy, electrostatic force microscopy, atomic force microscopy and scanning thermal microscopy (Bartoov et al., 2002; Akaki et al., 2002; Rothery et al., 2003). Most of these techniques involve biochemical processing that requires specific equipment and may also alter the vitality of the sperm analysed.

The sperm cell is almost transparent in conventional bright field microscopy, since its optical proprieties differ slightly from the surrounding liquid, generating little contrast. However, a light beam that passes through a spermatozoon undergoes a phase change, in comparison to the surrounding medium, the amplitude of which depends on the light source, the thickness and the integral refractive index of the object itself. A qualitative visualization of this phase contrast may be obtained by contrast interference microscopy (phase contrast or Nomarski/Zernicke interferential contrast microscopy). Recently, efforts have been renewed to improve Differential Interferometric Contrast methods in order to provide quantitative information in microscopy (Kou et al., 2010; Bon et al., 2009). On the other hands over the last few years, DH has been established as a valid non-invasive, quantitative, label free, high resolution, phase-contrast imaging technique in microscopy. In fact, this technique has been successfully applied to image a variety of cell types (Carl et al., 2004; Marquet et al., 2005; Kemper et al., 2006; Charrière et al., 2006; Di Caprio et al.,2010) to obtain additional information about their structure. In a recent preliminary study in human spermatozoa using DH, a statistically significant difference in phase shift was observed when comparing normal sperm with oligoasthenozoospermic sperm (Crha et al., 2011), however the authors did not measure standard morphological parameters using this technique, nor investigate the characteristics of sperm with nuclear vacuoles. One of the main advantages of DH is that a high-resolution 3-D quantitative sample imaging can be automatically produced by numerical refocusing of a 2-D image at different object planes without any mechanical scanning (Dubois et al., 1999; Ferraro et al., 2005). Moreover, using a single acquired image it is possible both to reduce the size of the mass storage devices required for image saving and to achieve a fast image transfer.

In this study, we have compared semi-automated digitally enhanced Nomarski microscopy (DESA) with DH to study morphometrical, morphological and volumetrical measurements in normal and vacuolated human spermatozoa. In particular, it is shown that DH is a viable tool for measuring the head volume either in the presence as well as in absence of vacuoles.

## MATERIALS AND METHODS

### *Specimen collection and analytical procedures*

Ejaculates were collected by masturbation from 15 males scheduled to undergo IVF at the Centro Fecondazione Assistita (CFA), Naples. After liquefaction, 0.5-1.0 ml of each specimen was processed using a double-density-gradient centrifugation method (Percoll, Sydney IVF). The final pellets were resuspended in Ham's F-10 (Gibco) and used to prepare slides for DESA. Briefly, a semen aliquot containing $2x10^6$ spermatozoa was washed 2x by centrifugation with phosphate-buffered saline (PBS). Then the pellet was suspended in 100 μl of 2% formaldehyde in PBS and fixed for 10 min at room temperature. A 10 μl aliquot was then spotted onto a clean microscope slide, allowed to air dry and mounted with PBS:Glycerol (1:1) (v/v).

In order to compare same cells by both DESA and DH, a grid of 20x20 circles (with a radius of 100μm) was placed over the microscope. This grid was made by a photolithographic process that allows the transfer of the shape of the grid from a mask to a photo-sensitive polymer. At the end of this process, the whole surface of the slide was covered with a 1μm-thick photo-sensitive material except the circles. A 10 μl aliquot was than spotted on the grid and finally a cover slide was placed to seal all. Only cells that fell in the circles of the grid were analysed (Fig 1).

Slides were first examined under immersion oil using an inverted microscope (TI-DH; Nikon Instruments Italia) equipped with Nomarski optics enhanced by digital imaging to achieve a magnification of up to X 1500. The images were captured by a colour video camera for high-quality image production and  analysed using image processing software (NIS-Element Documentation, Nikon).

### *Principles of Digital Holgrapy*

In holography, an object is illuminated by a collimated, monochromatic, coherent light with a wavelength λ. The object scatters the incoming light forming a complex wavefield (the *object*

*beam*):

$$O\mathrm{x,y} = | O\mathrm{x,y} | \, e^{j\varphi(x,y)} \tag{1}$$

where $|O|$ is the amplitude and $\varphi$ the phase, $x$ and $y$ denote the Cartesian coordinates in the plane where the wavefield is recorded (*hologram plane*). The phase $\varphi(x,y)$ incorporates information about the topographic profile of the object under investigation because it is related to the optical path difference (OPD), which depends on the refractive index and height both of the biological sample and of the material containing the object itself :

$$\varphi(\mathrm{x,y}) = \frac{2\pi}{\lambda} \cdot \mathrm{OPD} \tag{2}$$

where a transmission configuration has been considered. The purpose of holography is to capture the complete wavefront, and in particular the phase $\varphi$ to obtain quantitative information about the topographic profile of the object (Cuche et al., 1999). Since all light sensitive sensors respond to intensity only, the phase is encoded in the intensity fringe pattern adding another coherent background wave $R(\mathrm{x,y}) = | R(\mathrm{x,y}) | e^{j\varphi(\mathrm{x,y})}$, called the *reference beam*. This beam and the object beam interfere at the surface of the recording device. The *hologram* is proportional to the intensity of this interference pattern. In Digital Holography the hologram is acquired by a CCD (or CMOS) camera array, i.e. a two-dimensional rectangular raster of $M \times N$ pixels, with pixel pitches $\Delta\mathrm{x}$ and $\Delta\mathrm{y}$ in the two directions.

***Image reconstruction***

The image reconstruction procedure allows the retrieval of a discrete version of the complex optical wavefront present on the surface of the object under test. This optical wavefront is obtained by a numerical back propagation of spatially filtered product between the acquired hologram and a numerical replica of the reference beam (Cuche et al., 1999; Coppola et al., 2004; Ferraro et al., 2004). Thus, the reconstruction procedure allows to simultaneously determine both the intensity and especially the  phase distribution $\varphi(m,n)$ of the optical wavefront of the specimen. Where $\varphi(m,n)$ is the discretized version of the phase distribution φ(x,y) and *m*,*n* are positive integer numbers that identify the *m*-th row and *n*-th column of the pixels matrix of the CCD (CMOS) camera. By inverting eq. (2) and considering an homogeneous material with refractive index $n_c$, from the reconstructed phase distribution, the thickness distribution *s(m n)* of the object under investigation can be obtained as follows:

$$\text{OPDm,n} = \frac{\lambda}{2\pi}\,\varphi\,(\text{m,n}) = \frac{\lambda}{2\pi}\arctan\frac{\text{Im}[Q\text{m,n}]}{\text{Re}[Q\text{m,n}]} \tag{3}$$

where ***Q(m,n)*** is the discrete version of the optical rectostruced wavefront on the object surface, ***Im*** and ***Re*** are the imaginary and real part of the reconstructed optical field, respectively. The relation between the OPD and the thickness of the cell is *OPD(m,n)=s(m,n)·($n_c$- $n_s$)* where $n_c$ is the refractive index of the cell and $n_s$ is the refractive index of the surrounding medium.

Finally, the possibility offered by Digital Holography to manage the phase of the reconstructed image allows the removal and/or compensation of any unwanted wave front variations, such as optical aberrations (spherical, coma, tilt) and slide deformations (Ferraro et al., 2003; Coppola et al., 2010).

*Statistical analysis*

Statistical analyses were carried out by Student's *t* test. Probability values lower than or equal to 0.05 were considered significant.

**RESULTS**

A total of 2000 digitilized sperm heads were analysed for six primary parameters (length, width, perimeter, area, number and size of vacuoles). DESA analysis revealed that the mean values for length, width, perimeter and area of the sperm head were $5.18 \pm 0.64\mu m$ (mean $\pm$ SD) (range: 3.63 to 7.87), $3.53 \pm 0.45\mu m$ (range: 2.37 to 5.62), $13.75 \pm 1.35\mu m$ (range: 9.64 to 19.68), $14.12 \pm 3.03\mu m^2$ (range: 7.77 to 25.82). In the fifteen ejaculates 62.8% of spermatozoa with one or more vacuoles were found. The number of vacuoles per sperm ranged from 0 to 8 (mean: $2.01 \pm 1.78$) measuring $0.03\text{-}5.90\mu m^2$ in area (Fig. 2).

We used DH to evaluate the morphology of 200 of the afore-mentioned spermatozoa (Fig. 3). Figure 3A is an example of an acquired hologram, with the frange pattern highlighted in the inset. In Fig. 3B a pseudocolor plot of the phase-contrast map reconstruction of a human spermatozoon is shown. The colour-bar shows the value in rad of the phase difference which depends on the optical density and thickness of the biological sample. Fig. 3C illustrates the quantitative reconstructed morphology obtained by applying eq. 3 to the phase-map contrast.

It is important to note that this three-dimensional image is obtained from the reconstruction of a single acquired hologram, without the use of any mechanical scanning, allowing us to carry out numerical analyses of the six primary sperm parameters mentioned above. No significant differences were observed in the gross morphometric values of the sperm cells analyzed (using DESA or DH; Table 1).

In Fig. 4A and Fig. 4B we show the quantitative profiles of a spermatozoon along the lines AA' and BB' illustrated in Fig.3A, respectively. These profiles show, point by point, a quantitative value of the phase shift due to spermatozoan structure. Quantitative phase shift information from DH allows us to calculate the volume/mass of the sperm head. In Fig. 5, for example, an isoline plot relative to different heights of the sample is displayed. For each region defined by the isolines, the occupied area and the relative volume has been numerically estimated. The analysis revealed that the mean value for volume in normal sperm is $8.03 \pm 0.72\mu m^3$.

In Fig. 6 we show a quantitative comparison between a control spermatozoon and a spermatozoon with vacuoles. Fig 6A and B illustrates the height profile along the major axis of the sperm head for the defect-free spermatozoon and the spermatozoon with vacuoles, respectively.  In Fig. 6E the two profiles are shown together  to stress the differences. It's worth to note that both the shapes and the

185 point by point value of the height are different. In particular, the spermatozoon with vacuoles has a distinct depression in the profile (see the arrow in the figure). The profile of the normal spermatozoon results higher than that of the spermatozoon with vacuoles, whereas their 2D dimensions (such as area, and axes length) are similar. The difference in height difference implies a volume difference between the normal spermatozoon and the spermatozoon with vacuoles.

190 Table 2 shows three distinct groups of spermatozoa defined using two morphometric variables, head length and head width. The mean values of the volume for the three subpopulations were $5.76\pm073$ $\mu m^3$ (for length $< 2.9\mu m$ and width $< 4.2\mu m$), $8.24\pm0.78$ $\mu m^3$ (for $2.9 < length < 3.7\mu m$ and $4.2 < width < 5.3\mu m$), $10.13\pm0.81$ $\mu m^3$ (for length $> 3.7\mu m$ and width $> 5.3\mu m$). Mean values of the total volume of the spermatozoa minus the vacuoles volume are also reported.

## DISCUSSION

Here we have used DH as a novel approach for a more advanced morphological analysis of human spermatozoa, in particular to measure the head volume in the presence and absence of vacuoles.

In human sperm, the presence of vacuoles has been related to poor outcome in ART (Berkovitz et al., 2006), an increase in DNA fragmentation (Franco et al., 2008; Wilding et al., 2010) and abnormal chromatin packaging (Franco et al., 2011).

Nuclear vacuoles have been described as a crater defect in the spermatozoa of stallion (Johnson and Hurtgen, 1985), as a pouch formation (Bane and Nicander, 1965), a diadem defect (Blom, 1977) or a nuclear sperm defect (Miller et al., 1982) in bull spermatozoa, and as a crater defect (Johnson and Truitt-Gibert, 1982) or a pouch formation (Bane and Nicander, 1965) in boar spermatozoa. The defect is believed to originate during spermiogenesis as vacuoles have been found in both early and late spermatids (Johnson and Truitt-Gibert 1982). Nuclear vacuoles were shown by electron microscopy to be narrow-mouthed invaginations of the nuclear membrane into the nucleoplasm often filled with an amorphous, cytoplasmic material (Barth 1989). The predominant locations of the vacuoles are the apical region and the acrosome-postacrosomal sheath junction but they have also been found throughout the sperm head (Barth 1989).

Our results show that spermatozoa with vacuoles had a reduced volume probably due to variation of the inner structure of the sperm head with loss of material. We suggest that vacuolated spermatozoa with normal length and width (Bartoov et al., 2002) should be avoided for selection during the ICSI or IMSI procedure, until we acquire more data on the integrity and volume normal viable spermatozoon.

Recently, we have employed a microfluidic-system with DH on unstained bovine spermatozoa in their natural physiological surroundings (Di Caprio et al., 2010). This raises the possibility to use the same technique for a more complete analysis of human spermatozoa, with the additional possibility of sorting cells according to specific morphological criteria.

## TABLES

Table 1: Mean morphometric values of normal sperm heads  obtained by DESA and DHM techniques.

| Technique | Length ($\mu$m±SD) | Width ($\mu$m±SD) | Perimeter ($\mu$m±SD) | Area ($\mu$m$^2$±SD) | Volume ($\mu$m$^3$±SD) |
|-----------|---------|---------|-----------|---------|---------|
| DESA | 5.18±0.64 | 3.53±0.45 | 13.75±1.35 | 14.12±1.95 | - |
| DHM | 5.62±0.31 | 2.95±0.51 | 14.33±1.22 | 12.98±1.25 | 8.03±0.75 |

**Table 2:** Mean volumetric values of vacuolated sperm clustered in three different subpopulations

| Sperm dimensions | Volume ($\mu$m$^3$±SD) | |
|------------------|-------|-------|
| | Total | Total - Vacuoles |
| Length < 2.9$\mu$m<br>Width < 4.2$\mu$m | 5.76±0.73 | 3.99±0.76 |
| 2.9 < Length < 3.7$\mu$m<br>4.2 < Width < 5.3$\mu$m | 8.24±0.78 | 6.40±0.80 |
| Length > 3.7$\mu$m<br>Width > 5.3$\mu$m | 10.13±0.81 | 8.42±0.79 |

**FIGURE LEGENDS**

**Figure 1**. (A) A circle of a grid of 20x20 circles (with a radius of 100µm) placed over microscope slide. (B) Differential interference contrast micrograph of a sperm head. (C) Pseudocolor plot of the same spermatozoa.

**Figure 2**. High-power light microscope micrograph of sperm heads (1500X). (A) Crater-like appearance of nuclear vacuoles. (B,C,D) Spermatozoa with one or more vacuoles.

**Figure 3.** (a) Acquired hologram, a region is enhanced in order to show the interference pattern (inset). (b) Pseudocolor plot of a phase-contrast map for a human spermatozoon. (c) Pseudo 3-D representation of a human spermatozoon image reconstructed by DHM.

**Figure 4.** Profile plot along the lines (a) AA' and (b) BB', reported in Fig. 3(b).

**Figure 5.** Isolines plot of the reconstructed spermatozoon image.

 **Figure 6.** Comparison between a defect-free spermatozoa and a spermatozoa with vacuoles.

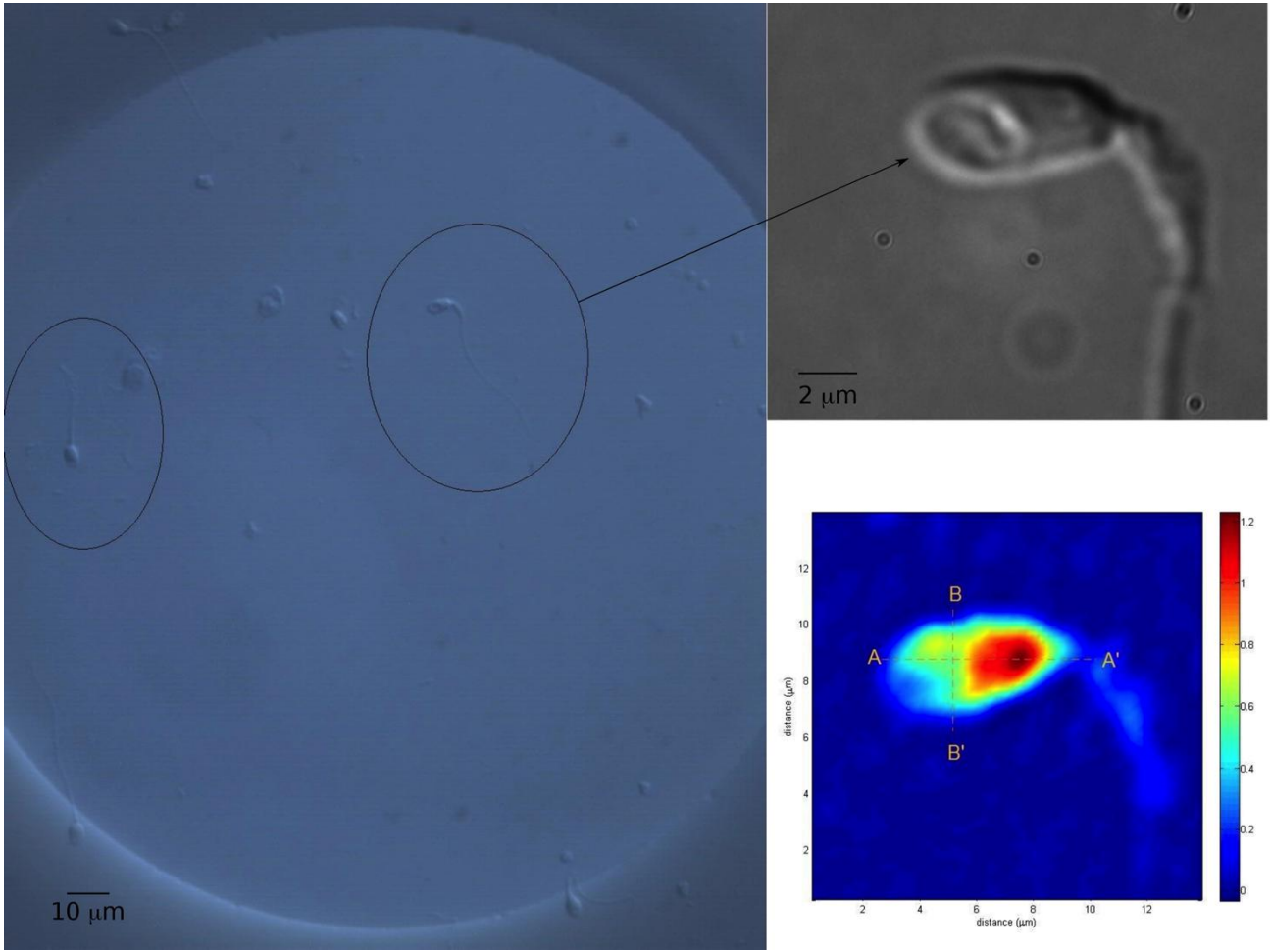

395

**FIG. 1**

400

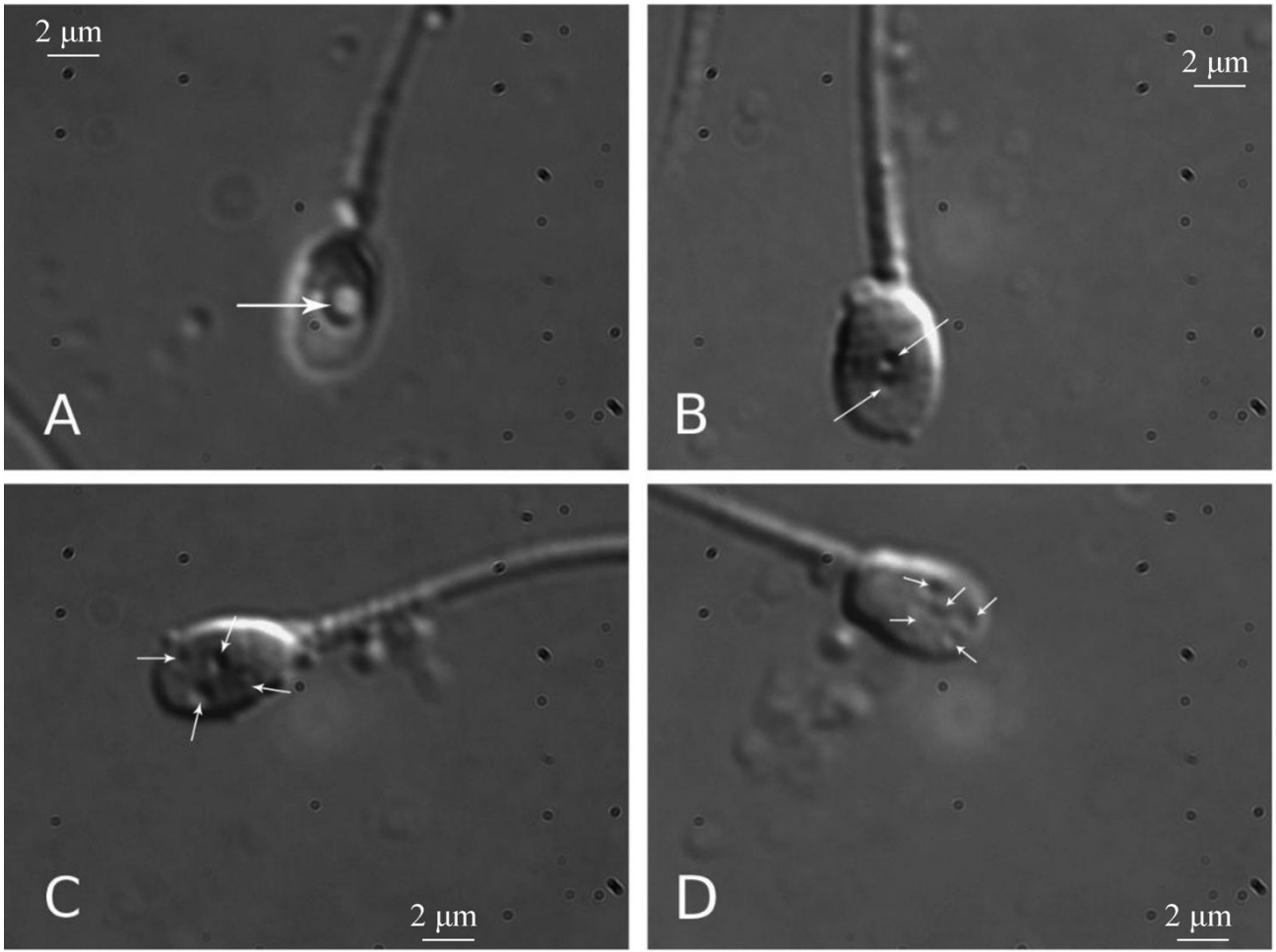

405    **FIG. 2**

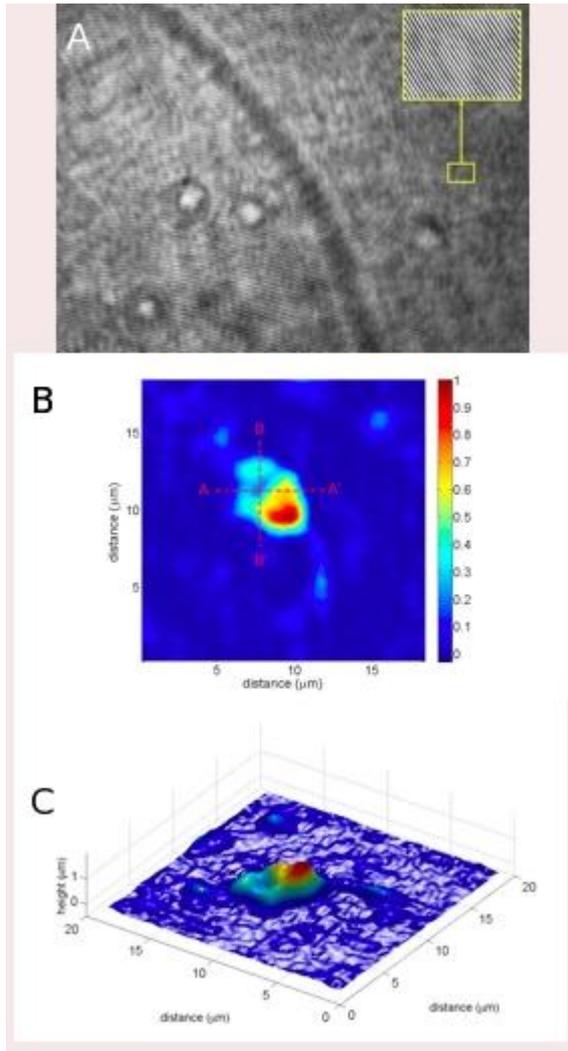

410

**FIG. 3**

415

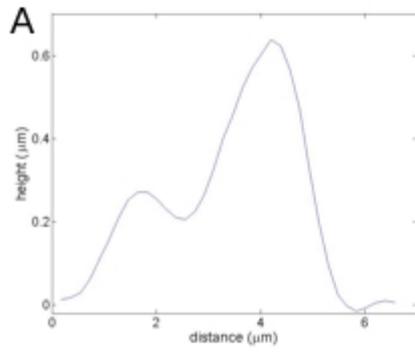 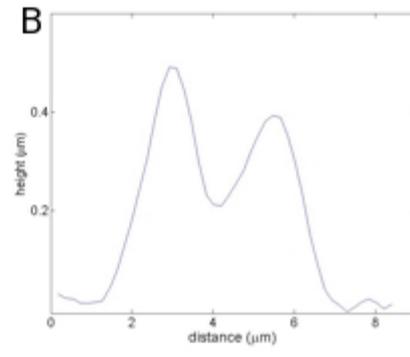

420    **FIG. 4**

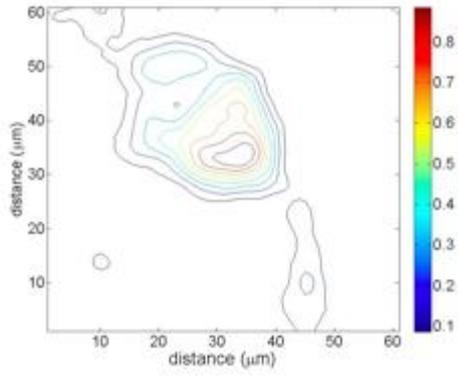

425    **FIG. 5**

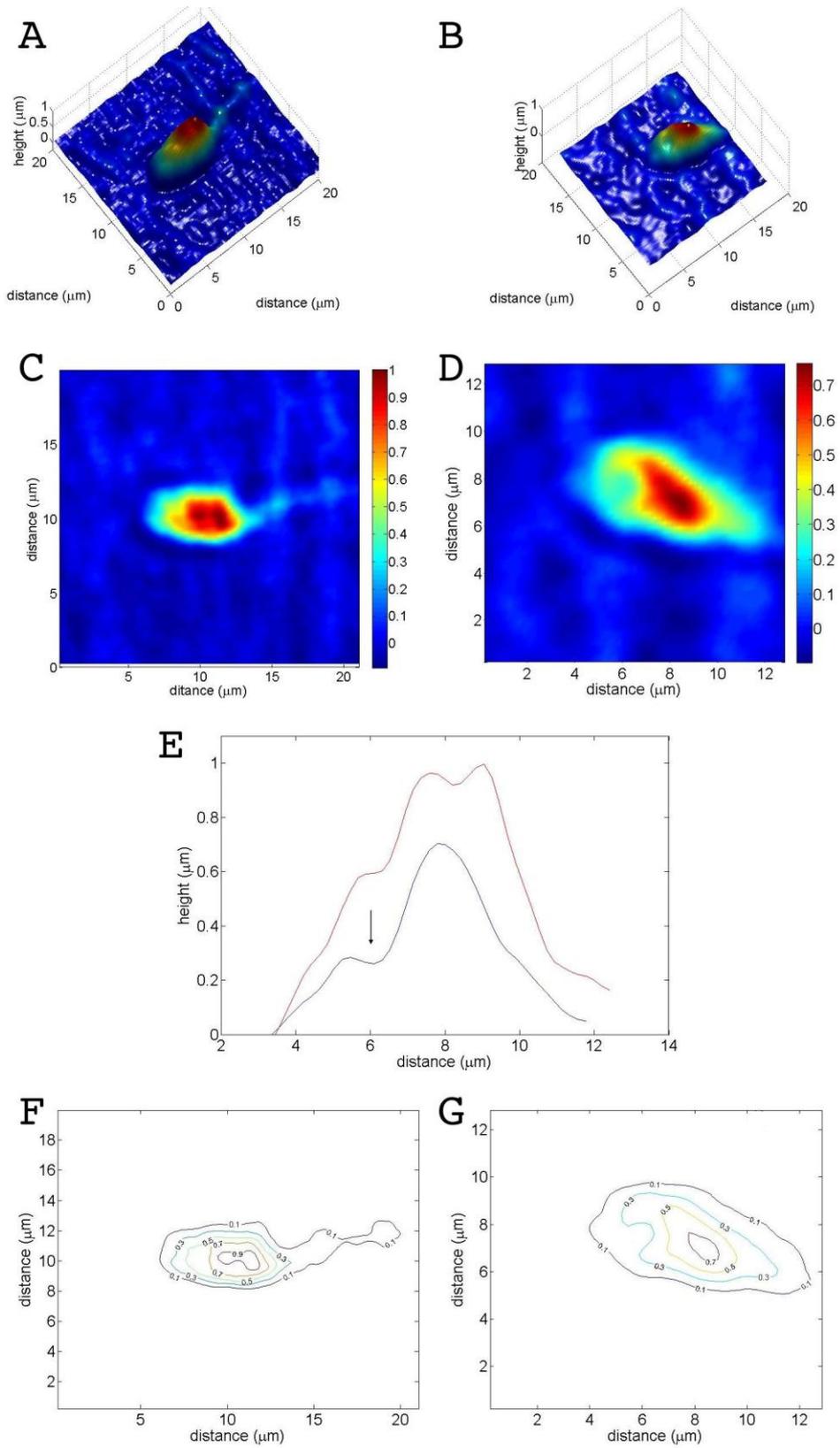

430

**FIG. 6**